\documentclass [12pt,a4paper,draft]{article}

\usepackage{times}

\DeclareFontFamily{OT1}{times}{}
\DeclareFontShape {OT1}{times}{m }{n }{ <-> ptmr }{}
\DeclareFontShape {OT1}{times}{bx}{n }{ <-> ptmb }{}
\DeclareFontShape {OT1}{times}{m }{it}{ <-> ptmri}{}
\DeclareFontShape {OT1}{times}{bx}{it}{ <-> ptmbi}{}
\usepackage{amsmath}
\usepackage{amsfonts}
\usepackage{amssymb}
\usepackage{latexsym}
\newcommand{\cl}{C \kern -0.1em \ell} % Clifford algebra
\newcommand{\CON}{\overline}          % quaternion (vector)conjugate

\newcommand{\Scal}{\mathbb{S}}        % scalar part unary  S[AB] = s
\newcommand{\Vect}{\mathbb{V}}        % vector part unary  V[AB] = v
\newcommand{\VEC}{\vec{\kern +.1em[}} % VEC-tor left  bracket   |>.. 
\newcommand{\TOR}{\vec{\kern +.2em]}} % vec-TOR right bracket   ..>| 
\newcommand{\BRA}{\langle\kern -.2em\langle} % Dirac BRA      <<...| 
\newcommand{\KET}{\rangle\kern -.2em\rangle} % Dirac KET      |...>> 

\newcommand{\Q}{[\hspace{1.mm}]} % quaternion operand A{\Q}B -> A[]B
\newcommand{\A}{(\hspace{.5mm})} % argument of function F{\A} -> F()

\newcommand{\INV}{{-1}}               % inverse function  
\begin{document}

\title{\vspace{-3cm}{\bf What is spin?}  } %~\\ \emph{\large Submitted to Am. J. Phys.}}
%               ------------------------------------

\author{{\bf André Gsponer}\\
\emph{Independent Scientific Research Institute}\\ 
\emph{ Box 30, CH-1211 Geneva-12, Switzerland}\\
e-mail: isri@vtx.ch\\}
\date{ISRI-03-10.13 ~~ \today}

\maketitle

\begin{abstract}

This is a late answer to question \#79 by R.I.\ Khrapko, {\bf ``Does plane wave not carry a spin?,''} Am. J. Phys. {\bf 69}, 405 (2001), and a complement (on gauge invariance, massive spin 1 and $\tfrac{1}{2}$, and  massless spin 2 fields) to the paper by H.C.\ Ohanian, {\bf What is spin?}, Am. J. Phys. {\bf 54}, 500--505 (1985).  In particular, it is confirmed that ``spin'' is a classical quantity which can be calculated for any field using its definition, namely that it is  just the non-local part of the conserved angular momentum. This leads to explicit expressions which are bilinear in the fields and which agree with their standard ``quantum'' counterparts.

\end{abstract}

%\section{Introduction}
%=====================

The problem of defining and calculating the intrinsic ``spin'' carried by plane waves is a recurring question (see, e.g., \cite{OHANI1986-, KHRAP2001-}) even though it has been definitely clarified sixty years ago by J.\ Humblet who recognized that such a definition is only possible for a wave-packet, that is a superposition of plane waves confined to a bounded four-dimensional region of space-time \cite[p.596]{HUMBL1943-}. This concept is in agreement with standard text books which make clear that plane waves of infinite extent are just mathematical idealizations and that physical waves correspond to ``fields produced a finite time in the past (and so localized to a finite region of space)'' \cite[p.333]{JACKS1975-}, \cite[p.99]{ROHRL1965-}. More generally, this concept agrees with Julian Schwinger's fundamental idea of source theory, namely that ``a physical description refers only to the finite space-time region which is under the experimenter's control'' \cite[p.78]{SCHWI1969-}.  Finally, in practical calculations, this concept leads to a simple recipe, applicable in many cases, which consists of approximating a wave-packet by a plane wave provided it is assumed that all contributions coming from infinitely remote space-time points (i.e., from so-called ``surface terms'') are zero.

It is therefore not surprising that the paper of H.C.\ Ohanian \cite{OHANI1986-}, as well as both answers to R.I.\ Khrapko's question \cite{KHRAP2001-}, i.e., \cite{ALLEN2002-, YURCH2002-}, are essentially based on this concept.  These papers focus on the spin 1 electromagnetic field:  What about spin~$\tfrac{1}{2}$ electrons, massive spin 1 particles, and possibly higher spin fields such as spin~2 gravitation?  Is there a simple, classical, i.e., ``non-quantical'' approach showing that these fields have indeed spin $\tfrac{1}{2}$, 1, and 2 respectively?

In the case of massless spin 1 particles, i.e., the electromagnetic field, Ohanian showed that the angular momentum $\vec{M}$ of a wave-packet can be decomposed into two part, $\vec{M}(\vec{r})=\vec{L}(\vec{r})+\vec{S}$, where by use of Gauss's theorem one of the parts is found to be non-local, i.e., independent of the spacetime point $\vec{r}$. This part, $\vec{S}$, therefore corresponds to an intrinsic angular moment carried by the wave-packet as a whole.  Written in terms of the electric field $\vec{E}$ and the vector potential $\vec{A}$, the volume density of this ``spin'' angular momentum is $\tfrac{d\vec{S}}{dV} = \vec{E} \times \vec{A}$, while the energy density is  $\tfrac{dU}{dV} = \tfrac{1}{2}(\vec{E}\cdot\vec{E}+\vec{B}\cdot\vec{B})$ where $\vec{B}$ is the magnetic field.  The normalized spin content of a wave-packet is therefore
$$
 \frac{d\vec{S}}{dU} = \frac{ \vec{E} \times \vec{A} }
                      {\tfrac{1}{2}(\vec{E}\cdot\vec{E}+\vec{B}\cdot\vec{B})}
       \eqno(1)  
$$
which is essentially Humblet's result \cite{HUMBL1943-}, confirmed by Jackson \cite[p.333]{JACKS1975-}, Rohrlich \cite[p.101]{ROHRL1965-}, Ohanian \cite{OHANI1986-}, and many others.

The occurrence of the non-observable and gauge-dependent vector potential $\vec{A}$ in a measurable quantity deserves an explanation: The crucial step in the derivation of $(1)$ is to express $\vec{B}$ as $\vec{B} = \vec{\nabla}\times\vec{A}$ in the definition $\tfrac{d\vec{M}}{dV} = \vec{r} \times (\vec{E}\times\vec{B})$ of the total angular momentum density. This enables the use of the Gauss theorem to discard a surface term and to isolate the non-local term $\vec{E}\times\vec{A}$ where, therefore, $\vec{A}$ can be considered as an abbreviation for the operational expression $(\vec{\nabla}\times)^\INV \vec{B}$, \cite{ROHRL2003-}.  As for the non-invariance of the spin density $\vec{E} \times \vec{A}$ in a gauge transformation $\vec{A} \rightarrow \vec{A} + \vec{\nabla} \chi$, the answer is that by using Maxwell's equations and discarding surface terms the additional contribution to the total spin content of a wave-packet can be written either $\iiint d^3x~ \vec{E} \times \vec{\nabla}\,\chi$ or $-\iiint d^3x~ \frac{\partial\vec{A}}{\partial t} \times \vec{\nabla}\,\chi$.  This means that the definition $(1)$ is consistent with Maxwell's equations and the notion of a wave-packet if $\vec{E} = -\frac{\partial\vec{A}}{\partial t}$, that is, only if the scalar potential is such that $\vec{\nabla}\phi=0$.  This condition, which is equivalent to the statement $\phi=\phi(t)$, defines a gauge which by Maxwell's equations for a source-free electromagnetic field yields the constraint $\vec{\nabla}\cdot\vec{A}=0$.  Therefore, while the separation of the total angular-momentum density into spin and orbital parts is not gauge invariant, the definition of spin for a wave-packet implies that the gauge $\vec{\nabla}\phi=0$ must be used. This makes this definition unique because for a source-free electromagnetic field the gauge function $\chi$ is then restricted by the requirement $\vec{\nabla}^2\chi=0$, whose only solution which is regular everywhere is a constant, so that $\vec{\nabla}\chi=0$, and $(1)$ is then gauge invariant.

In the case of a massive spin 1 field one has to use Proca's equations instead of Maxwell's.  It turns out that the calculations made in the case of Maxwell's field  can be repeated with the difference that the vector fields $\vec{E}$ and $\vec{B}$, as well as the vector and scalar potentials $\vec{A}$ and $\phi$, are now complex. The normalized spin content of a Proca wave-packet is then
$$
 \frac{d\vec{S}}{dU} = \frac{\tfrac{1}{2} ( \vec{E}   \times \vec{A}^* +
                                      \vec{E}^* \times \vec{A} )}
             {\tfrac{1}{2}(\vec{E}\cdot\vec{E^*} + \vec{B}\cdot\vec{B^*}) +
              \tfrac{1}{2}      m^2  (\phi\phi^* + \vec{A}\cdot\vec{A^*})}
       \eqno(2)  
$$
which reduces to $(1)$ for a real field of mass $m=0$.

Turning to the case of a spin $\tfrac{1}{2}$ field one meets the difficulty that the Dirac field is not a ``classical'' field in the sense that it is usually not expressed (as the Maxwell and Proca fields) in terms of scalars and three-dimensional vectors.  This is the reason why in the paper of Ohanian there is no explicit calculation showing that the spin of Dirac's field for an electron is indeed $\tfrac{1}{2}$.  Fortunately, there is a formulation of Dirac's theory (available since 1929!) which enables such a calculations to be made in a straight forward manner, and which allows a direct comparison with the Maxwell and Proca fields.  In this formulation \cite{LANCZ1929-} the four complex components of the Dirac field correspond to a scalar $s$ and a three-dimensional vector $\vec{v}$, which are combined into a complex quaternion $D = s + \vec{v}$ obeying the Dirac-Lanczos equation 
$$
      \CON{\nabla} D =  m D^* i\vec{\nu}             \eqno(3)
$$
where $\nabla$ is the 4-dimensional gradient operator and $\vec{\nu}$ a unit vector.  This equation, which is fully equivalent to Dirac's equation, has many remarkable properties which stem from the fact that it is written in the smallest possible algebra in which the Dirac electron theory can be expressed \cite{NOTE1,GSPON2002-}.  In the present application, following the procedure which lead to equations $(1)$ and $(2)$, and without using any ``quantum mechanical'' operator or rule such as projecting onto a subspace or calculating a ``Hilbert space scalar-product,'' one directly finds that the normalized spin content of an electron wave-packet is given by
$$
 \frac{d\vec{S}}{dU} =   \frac{1}{2} ~
                 \frac{\Vect[ D i\vec{\nu} \CON{D}^* ]}
  {\tfrac{1}{2}\Scal[ (\tfrac{\partial}{\partial it}D) i\vec{\nu} \CON{D}^*
                 - D i\vec{\nu} (\tfrac{\partial}{\partial it} \CON{D}^*) ]}
       \eqno(4)  
$$
which has an overall factor $\tfrac{1}{2}$ explicitly showing that the electron field has spin~$\tfrac{1}{2}$. In $(3, 4)$ the operator $\CON{\A}$ means changing the sign of the vector part, i.e., $\CON{s + \vec{v}}=s - \vec{v}$, while  $\Scal\Q$ and $\Vect\Q$ mean taking the scalar, respectively vector, part of the quaternion expression within the square brackets. 

We see that there is a great similarity between expressions $(1,2)$ and $(4)$.  The denominator is always the energy density which is obtained from the scalar part of the Poynting 4-vector, while the numerator is the vector part of the spin angular momentum pseudo 4-vector, i.e., $D i\vec{\nu} \CON{D}^* $ in Lanczos's formulation of Dirac's theory \cite{GSPON1998-}.  Moreover, if $\tfrac{1}{2}i\vec{\nu}$ is interpreted as the ``spin operator,'' expression $(4)$ is exactly the same as the one that would be written in quantum theory  for the ratio between the spin and energy densities.

   In principle, the method used here can be generalized to any spin.  This requires to know the explicit form of the corresponding energy-momentum tensor from which the analogue of the Poynting vector can be derived and used to calculate the non-local part of the angular momentum.  Unfortunately, there is no unique definition for this tensor, and there is much debate even for the spin~2 case which corresponds to gravitational radiation.  Nevertheless, using Einstein's energy-momentum pseudotensor, standard text books show how to derive the average rates of loss of angular momentum and energy of a point mass emitting gravitational waves in terms of its reduced mass quadrupole moments $Q_{jk}$, e.g., \cite[p.357]{LANDA1975-} and \cite[p.994]{MISNE1973-}.  This leads to the expressions
$$
\frac{dS_j}{dt} = - \frac{2}{5} m^2 ~\epsilon_{jkl}<\ddot{Q}_{kn}\dddot{Q}_{nl}>
 ~~~~ ,  ~~~~
\frac{dU}{dt}   = - \frac{1}{5} m^2               <\dddot{Q}_{jk}\dddot{Q}_{jk}>
  \eqno(5)
$$
which after division by one another show that gravitational radiation has spin 2, as is confirmed by comparing with the corresponding expressions for the electromagnetic angular momentum and energy loss rates of a point charge of 4-velocity~$\dot{Z}$ and electric charge $e$, see, e.g., \cite[p.175,~p.206]{LANDA1975-}
$$
  \frac{dS_j}{dt} = - \frac{2}{3} e^2 ~\epsilon_{jkl}<\dot{Z}_{k}\ddot{Z}_{l} >
   ~~~~ ,  ~~~~
  \frac{dU}{dt}   = - \frac{2}{3} e^2               <\ddot{Z}_{j}\ddot{Z}_{j} >
  ~~.\eqno(6)
$$

In conclusion, the ``spin'' content of any field can be defined and calculated without any reference to quantum theory.  In particular, if a polarized plane wave is used to calculate expressions $(1, 2)$ or $(4)$ one obtains a result that is 1 or $\frac{1}{2}$ times a normalized unit of angular momentum. If the corresponding wave is attributed to a single quantum such as a photon, a weak interaction boson, or an electron, this unit can be taken as the measured value of $\hbar$.  However, in order to consistently deal with fields containing a single or a small number of quanta, the classical theory is not sufficient:  It must be supplemented by a quantum interpretation in which the fields themselves become  dynamical variables \cite[p.751]{JACKS1975-}.  Finally, it is clear that ``spin'' has nothing to do with a vortex or a whirl which would be carried by a wave or a wave-packet:  It is simply the non-local part of the angular momentum that derives from the dynamics implied by the wave-equations defining the field.

~~\\
\noindent{\bf \Large Acknowledgments}
%====================================

~\\ The author thanks  Prof.~J.D.~Jackson, Prof.~F.~Rohrlich, Dr.~J.-P.~Hurni, and Dr.~A.B.~van Oosten for valuable comments and suggestions.

\end{document}